\documentclass[dvips]{acta}
\usepackage{supertabular,lscape,epsfig}
\usepackage{amssymb}
\usepackage{amsmath}

\usepackage{amssymb} 
\usepackage{hyperref}

\SetPages{0}{0}

\SetVol{71}{2021}

\begin{document}

\begin{Titlepage}

\Title{Wide-orbit exoplanets are common.\\Analysis of nearly 20 years of OGLE microlensing survey data}

\Author{
R.~~P~o~l~e~s~k~i$^1$,~~
J.~~S~k~o~w~r~o~n$^1$,~~
P.~~M~r~{\'o}~z$^2$,~~
A.~~U~d~a~l~s~k~i$^1$,\\
M.\,K.~~S~z~y~m~a~{\'n}~s~k~i$^1$,~~
P.~~P~i~e~t~r~u~k~o~w~i~c~z$^1$,~~
K.~~U~l~a~c~z~y~k$^3$,~~
K.~~R~y~b~i~c~k~i$^1$,\\
P.~~I~w~a~n~e~k$^1$,~~
M.~~W~r~o~n~a$^1$,~~and~~
M.~~G~r~o~m~a~d~z~k~i$^1$
}{
$^1$Astronomical Observatory, University of Warsaw, Al.~Ujazdowskie~4, 00-478~Warszawa, Poland\\
e-mail: rpoleski@astrouw.edu.pl\\
$^2$Division of Physics, Mathematics, and Astronomy, California Institute of Technology, Pasadena, CA 91125, USA\\
$^3$Department of Physics, University of Warwick, Gibbet Hill Road, Coventry, CV4~7AL,~UK
}

\Received{Month Day, Year}
\end{Titlepage}

\Abstract{
We use nearly 20 years of photometry obtained by the OGLE survey to measure the occurrence rate of wide-orbit (or ice giant) microlensing planets, \ie with separations from $\approx5~\mathrm{AU}$ to $\approx15~\mathrm{AU}$ and mass-ratios from $10^{-4}$ to $0.033$. In a sample of 3112 events we find six previously known wide-orbit planets and a new microlensing planet or brown dwarf OGLE-2017-BLG-0114Lb, for which close and wide orbits are possible and close orbit is preferred. We run extensive simulations of the planet detection efficiency, robustly taking into account the finite-source effects. We find that the extrapolation of the previously measured rate of microlensing planets significantly underpredicts the number of wide-orbit planets. On average, every microlensing star hosts $1.4^{+0.9}_{-0.6}$ ice giant planets.
}{
gravitational microlensing --- extrasolar ice giants 
}

\section{Introduction} 

The formation and evolution of planetary systems can be understood well only if we can study different populations of planets. Among the extrasolar planets, the ice giants are particularly hard to detect and study. The formation of Uranus and Neptune cannot be explained in the standard core-accretion model due to too low surface density of protoplanetary disc and too short disc lifetime (Pollack \etal 1996). Hence, the migration was proposed to explain the observed properties of Uranus and Neptune (Thommes \etal 1999, Tsiganis \etal 2005) and explaining Uranus and Neptune formation is still a subject of active research (\eg Izidoro \etal 2015).

Solar System ice giants have orbital periods so long that efficient finding of their analogs around stars other than the Sun cannot be done using methods that depend on a periodic phenomenon such as radial velocity or transit methods (Kane 2011). Such planets cannot be detected using direct imaging in the foreseeable future. The planet detection technique that can be used to study exoplanet ice giants is gravitational microlensing. During a microlensing event, the light of the background source star is bent in the gravitational field of the foreground lens system. If the lens system contains planets, then their signatures can be found in the magnification curve if the light from the source passes near these planets (Gaudi 2012). The sensitivity of microlensing depends on the mass of the lens, not its orbital period or brightness, hence, ice giant exoplanets can be found using microlensing. 

The microlensing technique allows discovering not only wide-orbit planets but even more extreme objects: free-floating planets (Mr\'oz \etal 2017, 2020), which are gravitationally unbound to any star. The free-floating planets cause short-timescale events which show only signal from a single lens. The event timescale is proportional to the square root of the mass, hence, a short timescale indicates that the lens is a planetary-mass object. However, one can question the interpretation of the shortest-timescale events as free-floating planets because such events can be also caused by bound planets on wide orbits for which stellar hosts did not show up in the light curves (Ryu \etal 2013). One needs to know the abundance of bound planets, and most importantly wide-orbit planets, to constrain the fraction of free-floating planet candidates that are due to bound planets (Clanton and Gaudi 2017). Here, we measure for the first time the abundance of wide-orbit planets from microlensing data, in particular, we use data from the Optical Gravitational Lensing Experiment (OGLE) that has already led to discovery of a few wide-orbit planets. Previous statistical studies of microlensing planets (\eg Cassan \etal 2012, Suzuki \etal 2016) were not sensitive enough to allow constraining the planet rate at large separations and small mass-ratios accurately. 

The main parameters characterizing microlensing planets are: $q$ -- planet to star mass-ratio and $s$ -- projected planet-star separation relative to the Einstein ring radius $\theta_\mathrm{E}$ (Gould 2000):
$$
\theta_\mathrm{E} = \sqrt{\kappa \pi_\mathrm{rel} M };~~~~\kappa\equiv\frac{4G}{c^2\mathrm{AU}} = 8.1 \frac{\mathrm{mas}}{M_\odot},
$$
where $\pi_\mathrm{rel}$ is the relative lens-source parallax and $M$ is the lens mass. Here, as a wide-orbit planet we consider a lens component with $q<0.033$ and $s>2$. The mass-ratio limit is $10\%$ higher than typically assumed in microlensing ($q = 0.03$; Bond \etal 2004) because one of the events considered here (OGLE-2016-BLG-0263; Han \etal 2017$a$) has a borderline mass-ratio of $0.0306\pm0.0008$. We consider only planets with $q>10^{-4}$ due to very small sensitivity to planets with smaller mass-ratios.

For a typical Galactic microlensing event, $\theta_\mathrm{E}$ projected at the lens distance ($D_l$) is: $\theta_\mathrm{E}D_l \approx 2.5~\mathrm{AU}$. Our definition of wide-orbit planet $s>2$ translates to a projected separation larger than $\approx5~\mathrm{AU}$. The widest orbit planets considered here are $s=6$, which translates to $\approx15~\mathrm{AU}$. The microlensing planets are detected if the source crosses or passes close to a caustic: the curve on which a point-source magnification would be infinite. For $s>2$ there are two caustics: one next to the host star (central caustic) and one lying closer to the planet (planetary caustic). In this paper, we analyze only events that show signals from both planetary and central caustics. We do not consider planets that are uncovered only by source passing close to the central caustic because the central caustic signals are intrinsically harder to interpret. As an example, the event OGLE-2013-BLG-0911 was a central caustic anomaly with $q=0.032$ and $s=0.15$ or $s=6.8$ as shown in a detailed analysis by Miyazaki \etal (2020), but the same event was incorrectly interpreted as $q=2.6\times10^{-4}$ in the large-scale analysis by Shvartzvald \etal (2016). The proper analysis of central caustic anomalies in the OGLE data is a major effort beyond the scope of the present paper. We note that the planet detection efficiency of the OGLE 2002-2008 data was studied previously by Snodgrass \etal (2004) and Tsapras \etal (2016).

In order to describe magnification curves of wide-orbit planet events let us first consider the simplest microlensing events, \ie the point-source point-lens events. The magnification $A$ in such events for epoch $t$ is given by Paczy\'nski (1986) equation:
$$
A = \frac{u^2+2}{u\sqrt{u^2+4}},
$$
where separation $u$ is
$$
u = \sqrt{u_0^2+\left(\frac{t-t_0}{t_\mathrm{E}}\right)^2} \eqno(1)
$$
and $u_0$ is the impact parameter relative to $\theta_\mathrm{E}$, $t_0$ is the epoch corresponding to the closest approach, and $t_\mathrm{E}$ is the Einstein timescale. The wide-orbit planets have planetary caustics located $s-1/s$ from the host (Han 2006), hence, the time separation between $t_0$ and planetary anomaly is slightly larger than $t_\mathrm{E}\left(s-1/s\right)$. At this time the magnification caused by the host is low enough that the peaks from the host and from the planet are essentially separate. This is best seen for OGLE-2008-BLG-092LAb (Poleski \etal 2014), which is the widest-orbit microlensing planet known ($s = 5.26\pm0.11$).

In the next section, we describe our data and their reduction. Section~3 describes selection of microlensing events. Sections~4 and 5 present a search for anomalies and a calculation of detection efficiency, respectively. The calculation of the wide-orbit planet occurrence rate is described in Section~6. The summary and discussion are presented in Section~7.

\section{Observations and Data Reduction} 

We analyze data from the third and the fourth phases of the OGLE survey (OGLE-III and OGLE-IV, respectively). The OGLE survey operates 1.3~m Warsaw Telescope, which is situated at the Las Campanas Observatory (Chile). During the OGLE-III, the telescope was equipped with an eight chip CCD camera and this phase was conducted between 2001 and 2009 (Udalski 2003, Udalski \etal 2008; $\mathrm{HJD'}\equiv\mathrm{HJD}-2,450,000$ from 2073 to 4956). In 2010, a 32-chip CCD camera was installed, which marked the beginning of the OGLE-IV (Udalski \etal 2015). Here, we analyze the OGLE-IV data taken from 2010 ($\mathrm{HJD'}=5377$) when the OGLE-IV camera started regular observations until the end of 2019 ($\mathrm{HJD'}=8788$). Both OGLE-III and OGLE-IV cameras have the pixel scale of $0.26''$ and 2k$\times$4k CCD chips, which results in fields of view of $0.35~\mathrm{deg^2}$ and $1.4~\mathrm{deg^2}$, respectively. The OGLE survey uses the $I$-band filter for majority of observations and here we analyze the time-series photometry in the $I$ band only. The $V$-band observations are much less frequent and we use them only for source characterization. The $I$-band exposure times are between 100 and $120~\mathrm{s}$. The photometry is extracted using the Difference Image Analysis (DIA; Alard and Lupton 1998, Wo\'zniak 2000). The DIA reports underestimated uncertainties and we correct them using the phenomenological model (Skowron 2009, Skowron \etal 2016). We analyze the OGLE-III and OGLE-IV photometric time series separately.

We analyze the fields with at least 300 epochs, which limits the sample to 88 OGLE-III fields and 85 OGLE-IV fields. The largest number of epochs is 2,500 and 16,000 in the OGLE-III and OGLE-IV data, respectively. We give details of the analyzed fields in a table provided at the OGLE Internet archive (see Section~7).

\section{Event Selection} 

\subsection{Initial Selection} 

We select candidate events in a way similar to the first steps in Mr\'oz \etal (2017, 2019). We consider stars brighter than $18.5~\mathrm{mag}$ (nine OGLE-IV fields with cadence of $1~\mathrm{h}$ or better) or $18~\mathrm{mag}$ (all other fields). We place a moving window that is $720~\mathrm{d}$ long and calculate the mean baseline flux and its dispersion ($\sigma_\mathrm{base}$) after removing $5\sigma$ outliers. Then, we search for bumps in the light curves with at least three consecutive points that are at least $3\sigma_\mathrm{base}$ brighter than the baseline.  We calculate the $\chi^2$ of a constant brightness model outside the moving window ($\chi_\mathrm{out}^2$) and restrict the sample to the candidates which have relatively small scatter outside the window by imposing $\chi_\mathrm{out}^2 / \mathrm{d.o.f.} < 2.0$. The limiting value is increased for the brightest OGLE-III targets because the corrections to the photometric uncertainties applied to these data do not account for a larger scatter for stars almost as bright as CCD saturation. We also expect a larger scatter for the brightest stars because they have high chances of being intrinsically variable (Soszy\'nski \etal 2013). Specifically, the $\chi_\mathrm{out}^2 / \mathrm{d.o.f.}$ limit is $3.0$ for stars brighter than $14~\mathrm{mag}$ and linearly drops to $2.0$ at $15.5~\mathrm{mag}$. This results in increasing the final sample by only $0.7\%$. We require at least five epochs to be detected on difference images during the candidate bump ($n_\mathrm{DIA} \geq 5$). The detections on difference images are counted only if the centroid is within $0.5''$ of the star detected on the reference image. For each bump, we select epochs brighter than the baseline by at least $3\sigma_\mathrm{base}$, use these epochs to calculate $\chi_{3+} = \sum_i\left(\left(F_i-F_\mathrm{base}\right)/\sigma_i\right)$, and limit the sample to bumps with $\chi_{3+} > 32$. We also require the amplitude of the bump to be larger than $0.1~\mathrm{mag}$. This sample still contains objects with bumps produced by instrumental effects, which are correlated in most cases. We search for pairs of objects with bumps at the same epochs and calculate a ratio of the number of common bumps ($N_1$) to the number of bumps detected for at least one object ($N_2$) and remove the pairs with $N_1/N_2 > 0.4$. We identify and remove a small number of candidates with bumps that are artificially produced by a brightening of a nearby ($<2''$) star.

Two fields (BLG535 and BLG610) produce particularly large number of candidates (2047 and 2021, respectively) and their inspection reveals that the majority of them are produced by spurious photometry taken during a single night ($\mathrm{HJD'} = 6029.X$). We exclude this night and re-run the event selection in these two fields.

Part of the time-series data analyzed here was previously searched for microlensing events by Mr\'oz \etal (2017): 2010--2015 data in nine fields with the highest cadence ($20~\mathrm{min}$ or $1~\mathrm{h}$). For these data we reuse Cut 2 of Mr\'oz \etal (2017) without rejecting objects with multiple bumps. The event selection in these data differs from the event selection in other data in only two aspects: the length of input light curve and the length of moving window used ($360~\mathrm{d}$ \vs $720~\mathrm{d}$). For all other data (including 2016--2019 data in the same fields) we follow the procedure described above.

We visually inspect all 14,351 light curves selected in this way. We select events that we consider useful for planet detection efficiency calculations, \ie we remove instrumental artifacts, light curves for which it is not obvious if they show microlensing events (mostly too poorly sampled and low-amplitude), variable stars (mostly cataclysmic variables), and binary-lens microlensing events with either stellar mass-ratios ($q>0.1$) or anomalies near the event peak (some were previously published as planets, \eg OGLE-2013-BLG-0341; Gould \etal 2014). This left us with 7228 candidates. We additionally mark events with anomalies that can be caused by wide-orbit planets so that we can independently check anomalies selected in an automated way (Section 4). During the visual inspection we discovered the event OGLE-2016-BLG-1928, which is the shortest-timescale microlensing event currently known and a very strong candidate for a free-floating planet (Mr\'oz \etal 2020).

\subsection{Photometry Cleaning} 

The next step is to clean the photometry from the outlying data points and remove a linear trend in the baseline data. We do it in several steps so that we do not remove the anomalies of the wide-orbit planets. 

First, we clean the light curves from outlying points that are outside the $360~\mathrm{d}$ long window. We calculate the scatter of this data $\sigma_{360}$ and remove the points which are $>5\sigma_{360}$ of the mean and a point before and a point after are within $2\sigma_{360}$ of the mean. Then we fit the point-source point-lens Paczy\'nski (1986) model. We first fit models with rectilinear motion and next fit annual microlensing parallax models (An \etal 2002). For fits here and in the next steps (unless noted otherwise) we use the ensemble Monte-Carlo Markov chain sampler implemented by Foreman-Mackey \etal (2013; \textsc{EMCEE}) and evaluate the light curves using the \textsc{MulensModel} package (Poleski and Yee 2019). We apply a prior on $t_\mathrm{E}$ and for this purpose we use an empirical distribution derived by Mr\'oz \etal (2017). We also apply prior on the measured blending flux ($F_b$): $1$ for $F_b\geq0$ and $\exp\left(-F_b^2/\left(2\sigma_{20}^2\right)\right)$ for $F_b<0$, where $\sigma_{20}$ is flux corresponding to $20~\mathrm{mag}$. We calculate residuals of the maximum \textit{a posteriori} model and fit to them a model linear in time: $\delta F=at+b$. If the coefficient $a$ is significant at $5\sigma$ level, then we subtract this trend from the original data. We re-calculate the scatter $\sigma'_{360}$ and apply the cleaning as above but without the $360~\mathrm{d}$ long window and $4\sigma'_{360}$ limit instead of $5\sigma_{360}$ limit.

We use the fits from previous sub-section to reject 49 events which peak beyond the time range covered by a given light curve. We also reject 48 events for which the baseline object is bluer and brighter than the main sequence turnoff. These events most probably have sources in the Galactic disk, which prevents obtaining a reliable estimate of the angular source radius, which is needed to calculate the planet detection efficiency. This left us with 7131 events.

\subsection{Final Selection}  

We fit the remaining light curves once more. For parallax models we limit north and east components of the microlensing parallax vector ($\pi_{\mathrm{E},N}$ and $\pi_{\mathrm{E},E}$, respectively) to $(-0.5, 0.5)$ range. We accept parallax models if $t_\mathrm{E} > 25~\mathrm{d}$ and $\chi^2$ is smaller by at least 150 than for the non-parallax model. We narrow down the sample to events with: a) a well-measured $t_\mathrm{E}$, b) significant source flux ($F_s$), and c) positive or reasonably negative blending flux. These requirements are imposed by: a) $t_\mathrm{E}/\sigma(t_\mathrm{E}) > 10$, b) $F_s/(F_s+F_b) > 0.1$ or $F_s$ corresponds to brighter than $18~\mathrm{mag}$, and c) either $F_b \geq 0$ or $F_b < 0$ and $|F_b|$ corresponds to fainter than $18~\mathrm{mag}$. These constrains result in 3112 events which we use to search for planets. To calculate detection efficiency, we restrict this sample to events with $\chi^2 / \mathrm{d.o.f.} < 3.0$, which results in 3095 events. Among them, $82\%$ are OGLE-IV events. The OGLE survey announces the on-going microlensing events at the Early Warning System (EWS) website (Udalski 2003) and $87\%$ of our sample was previously announced on the EWS website. All events discussed below were announced on the EWS website and we use EWS IDs in the text. Details of all events are provided at the OGLE Internet archive.

\section{Anomaly Search} 

To derive the planet rate one should search for planetary signals in a blind way. Most of the events analyzed here were already announced on the EWS website and hence searched for planets. We know the planets found in this way, thus, our search can only mimic a blind survey and should use properly defined criteria. 
Previous population studies of microlensing planets were based on the $\Delta\chi^2$ criterion. In most of these studies, either all events were high-magnification (\eg Gould \etal 2010) or most of the sensitivity came from the high-magnification events (\eg Suzuki \etal 2016). For the high magnification events most of the planet sensitivity comes from the part of the light curve that is near the peak and is short (Griest and Safizadeh 1998). Here, we study anomalies that are in the wings of the light curves, hence, the criterion on the $\Delta\chi^2$ seems to be not enough. It is possible that noise on timescales of a few hours can cause signals that can mimic the wide-orbit planets. Therefore, we design selection criteria that reject most of such cases. 

We search for points that are at least $5\sigma$ brighter than the point-lens model and consider them as anomaly if they are in a continuous set of at least three points each brighter by $>2.5\sigma$ than the point-lens model. We also constrain the minimum amplitude of the anomaly based on the number of anomalous points: $0.2$, $0.15$, $0.1$, and $0.045~\mathrm{mag}$ for $3$, $4$, $5$, and $>5$ anomalous data points, respectively. We note that one of the planets in our sample (OGLE-2011-BLG-0173Lb; Poleski \etal 2018) had two-night long anomaly (26 epochs) and observed amplitude slightly below $0.05~\mathrm{mag}$. The anomaly has to happen in the time range $(t_0-6t_\mathrm{E}, t_0-t_\mathrm{E})$ or $(t_0+t_\mathrm{E}, t_0+6t_\mathrm{E})$. The final criterion is on the $\Delta\chi^2$. We carried out experiments with different light curves and concluded that the $\Delta\chi^2$ limit should be on the order of 300. The sample of the wide-orbit planets with anomalies on planetary caustics is small and we would like to include in the rate analysis as many of them as possible but one of them (OGLE-2016-BLG-0263; Han \etal 2017$a$) shows the $\Delta\chi^2$ of 220 in the OGLE data. Hence, we decided to apply two selection criteria: $\Delta\chi^2 > 300$ (default selection) and $\Delta\chi^2 > 200$ (extended selection). The $\Delta\chi^2$ used in recent studies of microlensing planet abundance varies from 100 (Suzuki \etal 2016, Tsapras \etal 2016) to 500 (Gould \etal 2010). See also discussion in Yee \etal (2013).

\MakeTable{l|r|r|l}{12.5cm}{Detected wide-orbit planets} 
{\hline
event ID & $s$ & $q$ & reference \\
\hline
\hline
OGLE-2008-BLG-092 & $5.26\pm0.11$ & $(2.41\pm0.45) \times 10^{-4}$ & Poleski \etal (2014) \\
\hline
OGLE-2011-BLG-0173 & $4.65\pm0.13$ & $(4.5\pm1.5) \times 10^{-4}$ & Poleski \etal (2018) \\ 
\hline
MOA-2012-BLG-006 & $4.405\pm0.069$ & $(1.650\pm0.055) \times 10^{-2}$ & Poleski \etal (2017) \\
OGLE-2012-BLG-0022 & & & \\
\hline
OGLE-2012-BLG-0838 & $2.153\pm0.029$ & $(3.95\pm0.33) \times 10^{-4}$ & Poleski \etal (2020) \\
\hline
MOA-2013-BLG-605 & $2.39\pm0.05$ & $(3.6\pm0.7) \times 10^{-4}$ & Sumi \etal (2016) \\
OGLE-2013-BLG-1835 & & & \\
\hline
OGLE-2016-BLG-0263$^a$ & $4.72\pm0.12$ & $(3.06\pm0.08) \times 10^{-2}$ & Han \etal (2017$a$) \\
MOA-2016-BLG-075 & & & \\
KMT-2016-BLG-1515 & & & \\
\hline
\multicolumn{4}{p{12cm}}{For each event we provide IDs from OGLE, Microlensing Observations in Astrophysics (MOA; Bond \etal 2001), and Korean Microlensing Telescope Network (KMTNet; Kim \etal 2018) surveys. The first ID is the one used in the referenced paper. We note that OGLE-2008-BLG-092 was in Tsapras \etal (2016) sample and MOA-2012-BLG-006 was in Suzuki \etal (2016) sample.} \\
\multicolumn{4}{p{12cm}}{$^a$ -- this planet is detected only with extended selection criteria.}
}

The search for anomalies revealed six known wide-orbit planets (see Table~1), four other anomalous events discussed below, and a few false-positives caused by instrumental effects. We also searched for anomalies by relaxing each of the criteria. No viable planet candidate was found in these additional searches, which shows that there is no strong dependence of our final results on the detection criteria assumed. We also compare the anomalies selected here with the ones marked during the visual inspection described before. The lists are the same which shows that our data cleaning and automated anomaly selection worked properly.

\subsection{OGLE-2003-BLG-126} 

This event was previously analyzed by Skowron \etal (2009). They found a binary lens model with $q=0.604$, thus, the event is clearly non-planetary.

\subsection{OGLE-2007-BLG-030} 

\begin{figure}[htb]
{\centering
\includegraphics[width=.96\textwidth]{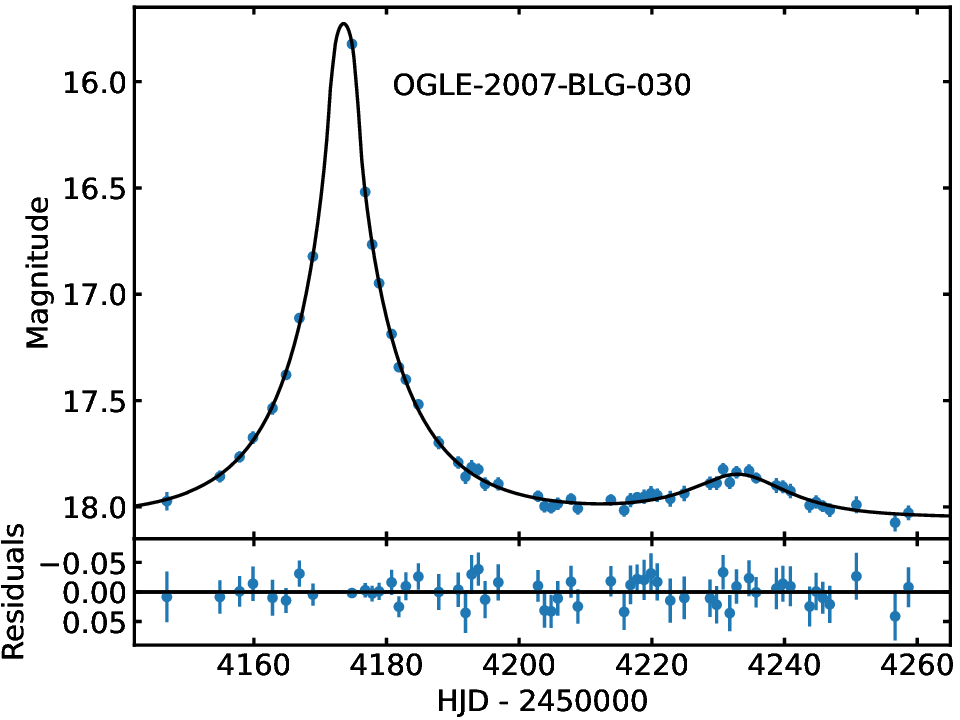} 

\bigskip
\includegraphics[width=.96\textwidth]{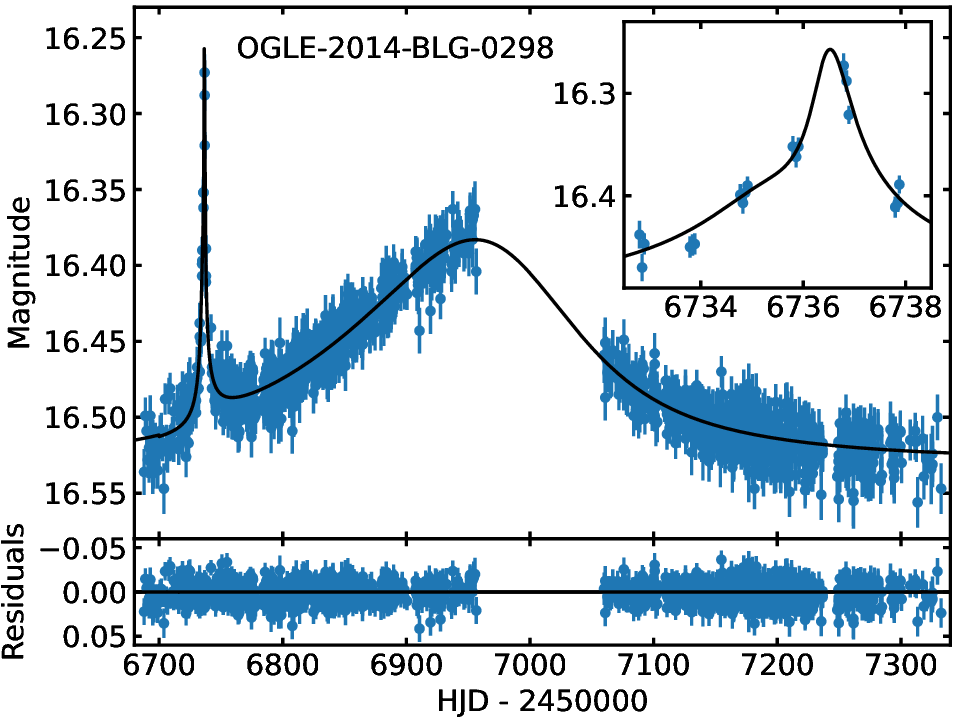}
}
\FigCap{Light curves of two astrophysical false positives. Top panel shows the event OGLE-2007-BLG-030 (maximum \textit{a posteriori} model with $s=3.75$ and $q=0.28$) and the bottom panel shows OGLE-2014-BLG-0298 ($s=0.32$ and $q=0.016$). The inset zooms-in on the anomaly.
}
\end{figure}

The event OGLE-2007-BLG-030 shows a low-amplitude anomaly after the main event -- see the light curve in the top panel of Figure~1. The length of the anomaly is relatively long, which suggests either a binary lens with non-planetary mass-ratio or a binary source. This event was not analyzed by Jaroszy\'nski \etal (2010) who searched for both the binary-source and the binary-lens events in the 2006--2008 OGLE-III data, most probably because of a low amplitude of the anomaly. We fit the binary-lens model and find a complicated posterior due to a poorly-sampled main peak. The median $q$ is $0.27$ and its $99\%$ lower limit is 0.14. The event is clearly not planetary, hence, we do not analyze it further.

\subsection{OGLE-2014-BLG-0298} 

The anomalous nature of this event was previously known but its detailed analysis was not published. The light curve is presented in the bottom panel of Figure~1. We fit the static binary lens with parallax and finite-source model and found that the close model ($s = 0.33\pm0.01$, $q = 0.016\pm0.002$) is preferred over the wide model ($s = 2.15\pm0.03$, $q = (4.4\pm0.5)\times10^{-4}$) by $\Delta\chi^2 = 56.4$. It is much more likely that the lens orbital motion is significant for the close model than for the wide model and including the orbital motion should further increase $\Delta\chi^2$. We conclude that this event is not a wide-orbit planet event.

\subsection{OGLE-2017-BLG-0114} 

\begin{figure}[htb]
{\centering
\includegraphics[width=\textwidth]{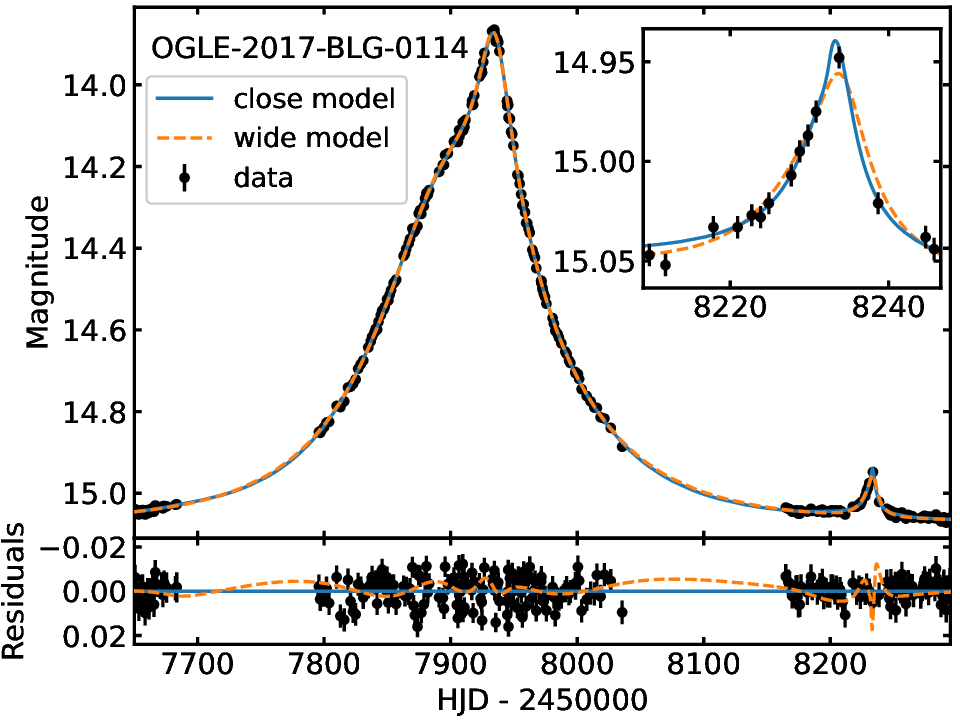}
}
\FigCap{Light curve of new planetary event OGLE-2017-BLG-0114. The solid blue line indicates the best close model ($s=0.30$) and the orange dashed line indicates the best wide model ($s=3.23$). The inset zooms-in on the anomaly.
}
\end{figure}

The event coordinates are $\mathrm{R.A.} = 17^\mathrm{h}21^\mathrm{m}57.74^\mathrm{s}$, $\mathrm{Dec.} = -29^\circ37'25.3''$ (Galactic coordinates $l=-3.38^{\circ}$, $b=3.96^{\circ}$). This falls in an OGLE field observed with a cadence of $1~\mathrm{d}$ and the event was not announced by either MOA or KMT surveys. The 2016 and 2017 data show a clear signal with an amplitude of $1.2~\mathrm{mag}$ -- see light curve in Figure~2. The shape of the signal near the peak is clearly asymmetric. The 2016 and 2017 photometry is not explained by the point-source point-lens model with microlensing parallax model. The timescale of the asymmetry is too short to be explained by the annual motion of the observer. This suggests that the source could be a member of a binary system which shows significant orbital motion during the microlensing event, or xallarap (Griest and Hu 1992). The 2018 data show additional shorter peak with an amplitude of $0.10~\mathrm{mag}$ (inset in Figure~2). We fit 2016-2018 data with a static binary lens model including xallarap and finite-source effects and consider both $s<1$ and $s>1$ solutions. The orbital period of the source shows multiple modes in a wide range from $140~\mathrm{d}$ to $500~\mathrm{d}$ and none of them is close to the orbital period of the Earth. We found the close solution ($s=0.2988\pm0.0087$, $q=0.0428^{+0.0055}_{-0.0084}$) to fit the data better than the wide solution ($s=3.23^{+0.18}_{-0.24}$, $q=0.0021^{+0.0035}_{-0.0007}$) by $\Delta\chi^2 = 45.0$. While this is not decisive, the close model seems much more likely. The mass-ratios of the two models differ significantly: the close model has a brown dwarf secondary while for the wide model the secondary is a planet. Both models are presented in Figure~2. One can see that they differ mostly during the anomaly (up to $0.02~\mathrm{mag}$) but for other parts of the light curve we see differences of up to $0.005~\mathrm{mag}$, which significantly contributes to $\Delta\chi^2$. Detailed analysis of OGLE-2017-BLG-0114 will be presented separately.

\section{Detection Efficiency} 

The binary-lens events are parameterized by at least six parameters: three are the same as for point-lens events, $s$, $q$, and $\alpha$ (an angle between the binary axis and the source trajectory). For most of the binary-lens events with planetary mass-ratios one also measures $\rho$ -- the radius of the source relative to $\theta_\mathrm{E}$. We aim to measure detection efficiency as a function of $s$ and $q$. Thus, there are two nuisance parameters: $\alpha$ and $\rho$. For $\alpha$ we assume a uniform distribution and marginalize over it (see below). The parameter $\rho$ is more problematic because of a few factors: it is not directly constrained for majority of the point-lens events (all except a small number of $u_0 \lesssim \rho$ events), its \textit{a priori} distribution is significantly asymmetric, and it affects the detection efficiency (Han \etal 2005). Hence, accounting for $\rho$ has to be carefully considered. Most previous detection efficiency studies used Galactic models to predict \textit{a prior} distribution of $\theta_\mathrm{E}$, estimated the angular radius of the source ($\theta_\star$) based on the source brightness and the brightness of red clump stars (Yoo \etal 2004), and combined the two: $\rho = \theta_\star/\theta_\mathrm{E}$. In previous studies, this equation was applied to three values of $\theta_\mathrm{E}$: $16^\mathrm{th}$, $50^\mathrm{th}$, and $84^\mathrm{th}$ percentiles. The detection efficiency was average of the values obtained for the three resulting $\rho$ values. In the following subsections, we first describe the source size estimates, then present the procedure for events fitted without parallax, and at the end discuss the modifications for parallax events.

\subsection{Source Size Estimates} 

The size of the source relative to $\theta_\mathrm{E}$ is estimated in two ways. First, events with the smallest $u_0$ and a well-sampled peak allow a direct estimate. We fit point-lens finite-source models and find useful $\rho$ constraints for 54 events. In some cases, these constraints are effectively upper limits but are useful nevertheless. The magnification for a point-lens finite-source model is calculated using methods by Gould (1994) and Lee \etal (2009) for $\rho < 0.1$ and $\rho > 0.1$, respectively.

Second, $\rho$ can be estimated indirectly by comparing the brightness of the source ($I_s$) with that of the red clump stars ($I_\mathrm{RC}$; both calibrated to the standard system) in the event vicinity. We assume that the amount of interstellar extinction towards source stars and nearby red clump stars is the same. The calculation also requires assuming the radius of red clump stars ($R_\mathrm{RC}$), source distance ($D_s$), and $\theta_\mathrm{E}$. Specifically:
$$
\theta_\star = \theta_\mathrm{RC} 10^\frac{I_s-I_\mathrm{RC}}{-5}, \eqno(2)
$$
where the angular source radius of the red clump stars is:
$$
\theta_\mathrm{RC} = \frac{R_\mathrm{RC}}{D_s}. \eqno(3)
$$
We assume $R_\mathrm{RC}$ of $11~R_\odot$, which is a typical red clump star radius as measured using the long-baseline interferometry (Gallenne \etal 2018).

The red clump properties in the OGLE-III fields were previously studied by Nataf \etal (2013) who measured $I_\mathrm{RC}$ on a sky grid with varying size. We use the Nataf \etal (2013) results if the nearest grid point is within $1'$ from the event. For other events we select stars within $2'$ and measure $I_\mathrm{RC}$ using the method described by Nataf \etal (2013). We use magnitudes calibrated to the standard system from either OGLE-III (Szyma\'nski \etal 2011) or OGLE-IV (proprietary data). The model fitted to the distribution of $I$ magnitudes has only five parameters: $I_\mathrm{RC}$, $\sigma\left(I_\mathrm{RC}\right)$, number of the red clump stars, and two parameters that describe the brightness function of the red giant stars. Following Nataf \etal (2013), the model also includes the red giant branch bump and the asymptotic giant branch bump. The parameters of their brightness distribution are defined relative to the parameters of the red clump.

To estimate the probability distributions of $\theta_\mathrm{E}$ and $D_s$ we run the simulation of microlensing in the Galactic model. We use the model by Clanton and Gaudi (2014) as modified by Poleski \etal (2020) and run it separately for equatorial coordinates of each event. The simulated events are then constrained by the posterior distributions of $t_\mathrm{E}$ and $\pi_\mathrm{E}$ (if it is measured). From the resulting parameter distributions, we obtain ten samples of $\theta_\mathrm{E}$ (denoted $\theta_{\mathrm{E},j}$) and their weights ($\sum_{j=1}^{10}w_j=1$) using the importance sampling technique, which allows efficient sampling of a distribution using small number of samples. Specifically, we draw ten samples from a normal distribution which has $\sigma$ larger by 10\% than the actual distribution and assign weights which are proportional to the ratio of the actual distribution to the normal distribution. We also measure parameters of $D_s$: median and $16\%$ and $84\%$ quantiles (which define $\sigma_-$ and $\sigma_+$). We randomly draw ten samples of $D_s$ from a two-piece normal distribution with $\sigma_-$ and $\sigma_+$ increased by 15\% in order to account for intrinsic scatter of the red clump stars radii and uncertainties possibly not taken into account.

Finally, we calculate $\rho_j = \frac{\theta_{\star,j}}{\theta_{\mathrm{E},j}}$.

\subsection{Non-parallax Events} 

We calculate detection efficiency on a grid of $s$ values from $2.0$ to $6.0$ with a step of $0.2$ and on a grid of $q$ from $10^{-4}$ to $0.033$ with 20 log-uniformly spaced values. To reduce the computational time, we consider the ranges of $\alpha$ that correspond to trajectories passing close to the planetary caustic. There are two ranges of $\alpha$ that correspond to two arcs of a circle. These $\alpha$ values are selected so that they pass planetary caustic at a distance that is larger of $\rho$ or 4.0 scaled to the Einstein ring radius of the planet ($\theta_\mathrm{E, pl}$), \ie:
$$
\alpha = \arcsin\left(\frac{u_0\pm\mathrm{max}\left(\rho,4\sqrt{\frac{q}{1+q}}\right)}{s-\frac{1}{s}}\right), \eqno(4)
$$
where $\sqrt{q/(1+q)}$ is a factor for scaling from $\theta_\mathrm{E, pl}$ to $\theta_\mathrm{E}$. We note that for a small sample of events with $u_0 > 1.5$, there is no sensitivity for detecting planets with the smallest $s$ values considered here, because the planetary and central caustics are separated by $s-1/s = 1.5 < u_0$. The Equation~(4) defines one range of $\alpha$ values to be checked and the other one is simply obtained by substituting $\alpha \rightarrow \pi - \alpha$. We set the grid spacing of $\alpha$ (denoted $\delta\alpha$) so that $\delta\alpha/\alpha$ is at least ten times denser than $\delta s/s$ and there are at least 60 values of $\alpha$. For each $\alpha$ value we check detectability of the planet. If a planet is detectable for $\alpha$ at the edge of the range, then the range is iteratively extended. 

To check if a planet would be detected for given $s$, $q$, $\alpha$, and $\rho$ we use the method by Rhie \etal (2000), which for point-lens parameters takes into account only maximum \textit{a posteriori} model, not the full posterior distribution of parameters. This method is appropriate for planetary caustic anomalies considered here, because single-lens model parameters $(t_0, u_0, t_\mathrm{E}, F_s, F_b)$ are constrained by data other than the candidate anomaly data. In the Rhie \etal (2000) method, one calculates point-lens ($A_{\mathrm{1L}, i}$) and binary-lens ($A_{\mathrm{2L}, i}$) magnification for each epoch $t_i$. For the binary-lens magnification calculation we use the \textsc{VBBL} algorithm (Bozza 2010, Bozza \etal 2018). The \textsc{VBBL} is initially applied only to epochs close to the expected anomaly times, \ie the range $(t_{0,\mathrm{pl}}-t_\mathrm{E,pl}, t_{0,\mathrm{pl}}+t_\mathrm{E,pl})$, where: 
$$
t_{0,\mathrm{pl}} = t_0 + \frac{t_\mathrm{E}}{1+q}\left(s-\frac{1}{s}\right)\cos\alpha
$$
and (Khakpash \etal 2019):
$$
t_\mathrm{E,pl} = t_\mathrm{E}\sqrt{q+\rho^2}.
$$
Next, we calculate fluxes for each epoch: $F_{\mathrm{1L}, i} = F_sA_{\mathrm{1L}, i}+F_b$ and $F_{\mathrm{2L}, i} = F_sA_{\mathrm{2L}, i}+F_b$. We then iteratively extend the time range for which the \textsc{VBBL} method is used if $(F_{\mathrm{2L}, i} - F_{\mathrm{1L}, i}) / F_{\mathrm{1L}, i} > \sigma(F_i)/\left(3F_i\right)$. Next, we scale the uncertainty of flux for each epoch assuming it is limited by the Poisson statistic: $\sigma(F_{\mathrm{2L}, i}) = \sigma(F_i)\sqrt{F_{\mathrm{2L}, i}/F_{\mathrm{1L}, i}}$. We apply the same criteria as in Section~4. Specifically, to calculate the $\chi^2$ for a single-lens model we use \textsc{scipy} (Virtanen \etal 2020) implementation of the Broyden-Fletcher-Goldfarb-Shanno algorithm, which finds maximum-likelihood parameters based on the $\chi^2$ and its gradient. In our fits, just a few function evaluations are enough to find the maximum likelihood model. Finally, the detection efficiency for event $i$ and parameters $s$, $q$, and $\rho_j$ (denoted $S_i(s, q; \rho_j)$) is the ratio of the number of grid angles for which a planet would be detected to the total number of grid angles in $2\pi$.

\subsection{Parallax Events} 

The procedure above assumes that the relative lens-source proper motion is rectilinear, which allows using Equation~(4) to find $\alpha$ ranges. This assumption is not valid for events with detectable parallax. Hence, the detection efficiency calculation procedure has to be modified for parallax events. For each epoch $t$ we calculate the normalized lens-source separation $u$. Then we invert Equation~(1) to calculate $t'$, which would result in the same $u$ value for a model which has the same $t_0$, $u_0$, and $t_\mathrm{E}$ but no parallax:
$$
t' = t_0 \pm t_\mathrm{E}\sqrt{u^2-u_0^2}, \eqno(5)
$$
where plus and minus signs correspond to $t>t_0$ and $t<t_0$, respectively. We illustrate this equation in Figure~3. For anomalies on planetary caustics, the detection efficiency depends on cadence and photometric uncertainties. The absolute differences $\left|t - t'\right|$ for consecutive epochs are comparable even if these differences are significant. Hence, the calculated $S_i(s, q;\rho_j)$ values are correct and we significantly simplify the calculations for parallax events.

\begin{figure}[htb]
{\centering
\includegraphics[width=\textwidth]{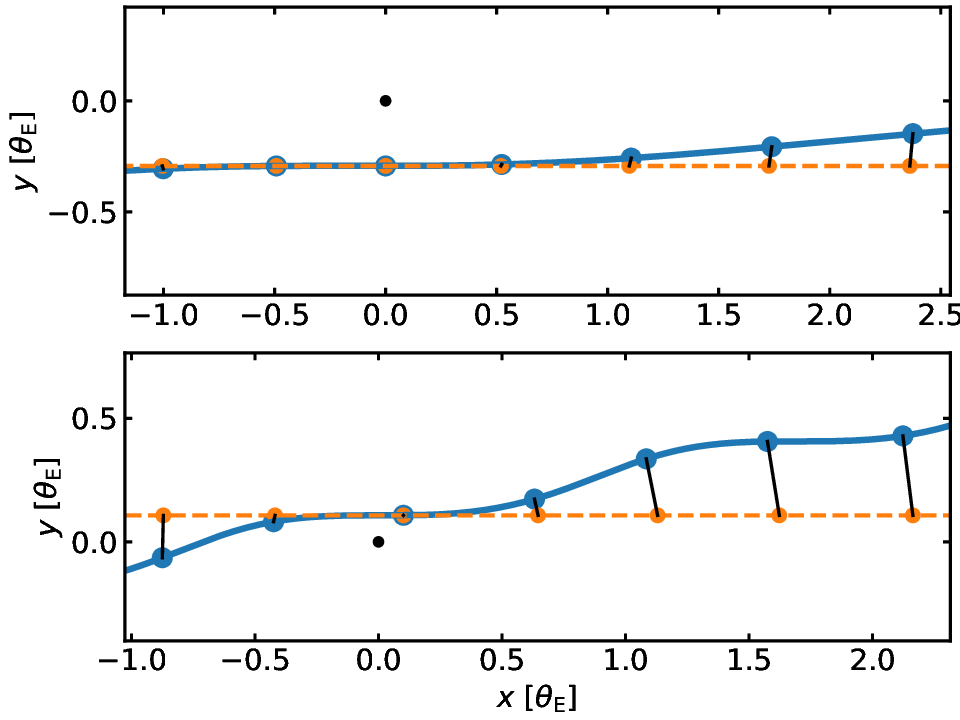}
}
\FigCap{Illustration of Equation~(5) for two example parallax events. In each panel, the solid blue line shows actual source trajectory relative to lens (black dot). Blue dots are uniformly spaced and are connected using black lines to the corresponding positions on the orange dashed line according to Equation~(5). The orange dashed line has the same $t_0$, $u_0$, and $t_\mathrm{E}$ as the parallax trajectory, but no parallax. The top panel shows OGLE-2014-BLG-0115 ($u_0=-0.29$, $t_\mathrm{E}=99.4~\mathrm{d}$, $\pi_{\mathrm{E}, E} = 0.09$, $\pi_{\mathrm{E}, N}=0.03$). The bottom panel shows OGLE-2015-BLG-0142 ($u_0=0.11$, $t_\mathrm{E}=195.8~\mathrm{d}$, $\pi_{\mathrm{E}, E} = 0.03$, $\pi_{\mathrm{E}, N}=-0.05$).
}
\end{figure}

\subsection{Results} 

Examples of detection efficiency calculations are presented in Figures~4 and 5. After calculating detection efficiency for a given event and ten $\rho_j$ values, we weight the results:
$$
S_i(s, q) = \sum_j^{10} w_j S_i(s, q; \rho_j).
$$
Then we sum the $S_i(s, q)$ for all events and obtain survey detection efficiency:
$$
S(s, q) = \sum_i S_i(s, q).
$$

\begin{figure}[htb]
{\centering
\includegraphics[width=\textwidth]{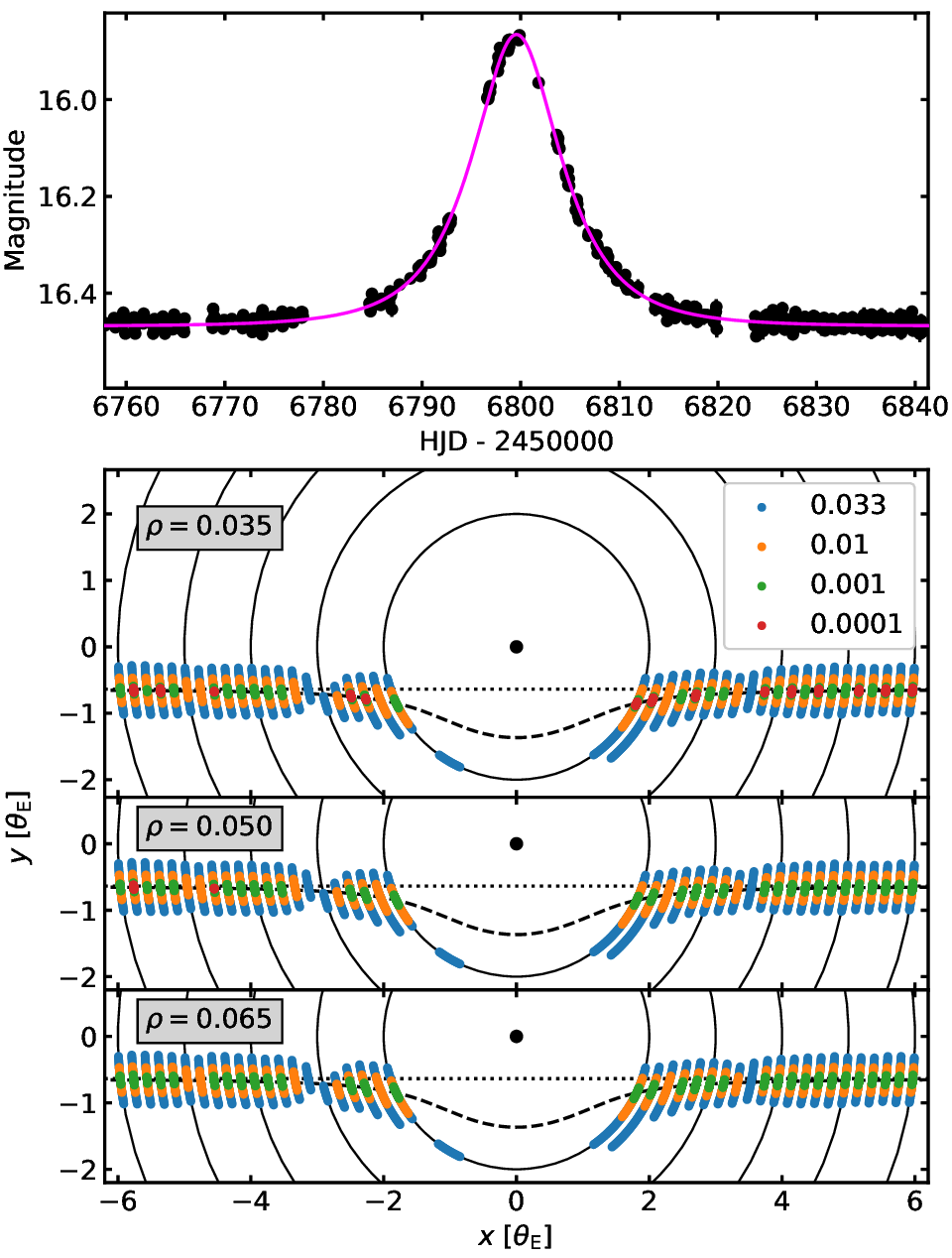}
}
\FigCap{Example detection efficiency calculations for event OGLE-2014-BLG-0729. The top panel shows the event light curve with fitted model in magenta ($u_0=0.64$, $t_\mathrm{E} = 6.9~\mathrm{d}$). The bottom panels show detectable positions of the planet -- dots with color-coded four fiducial values of $q$ (see legend). Each panel corresponds to a different value of $\rho$ as indicated in the top left corner. The black circle indicates the lens position and the dotted line is the source trajectory. The dashed line indicates the trajectory of the major image in the single-lens model (Paczy\'nski 1986). Solid circles are plotted at integer values of $s$ to guide the eye.
}
\end{figure}

\begin{figure}[htb]
{\centering
\includegraphics[width=\textwidth]{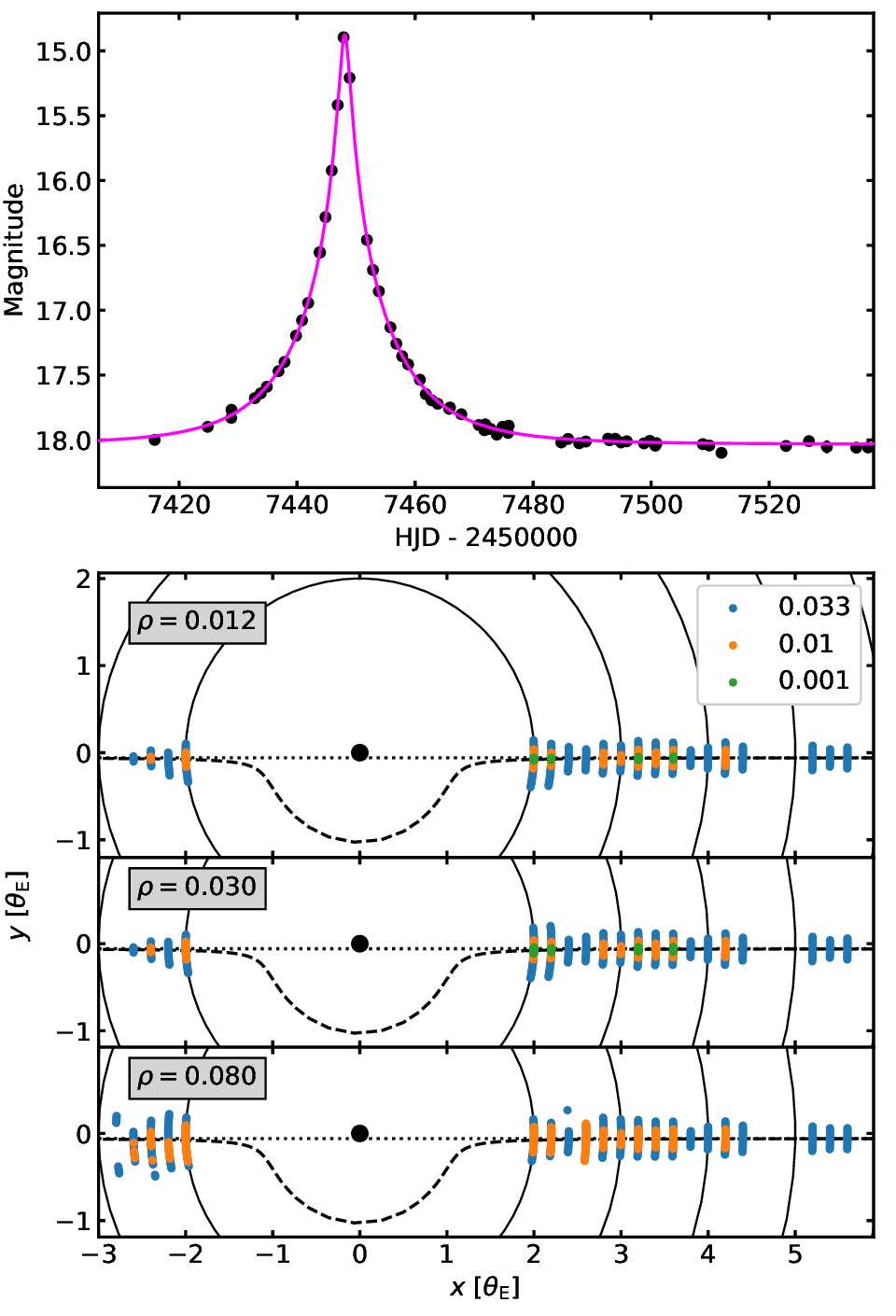}
}
\FigCap{Same as Figure~4 but for OGLE-2016-BLG-0094 ($u_0=0.058$, $t_\mathrm{E} = 15.6~\mathrm{d}$). The cadence of the photometry is lower and there is no detection efficiency at $q=10^{-4}$.
}
\end{figure}

We show the calculated detection efficiency for the default selection criteria in Figure~6. As expected, the detection efficiency increases for increasing $q$ and decreasing $s$. The extended selection criteria lead to $S(s, q)$ higher by on average $6.3\%$ than the default criteria. The survey sensitivity derived here is higher than that of Suzuki \etal (2016; their Figure~6) by a factor of a few at the largest separations considered here. At $s\approx2$ and small $q$ both surveys have similar sensitivity and ours is less sensitive for increasing $q$ (because we intentionally omit central caustic anomalies in this analysis). 

\begin{figure}[htb]
{\centering
\includegraphics[width=\textwidth]{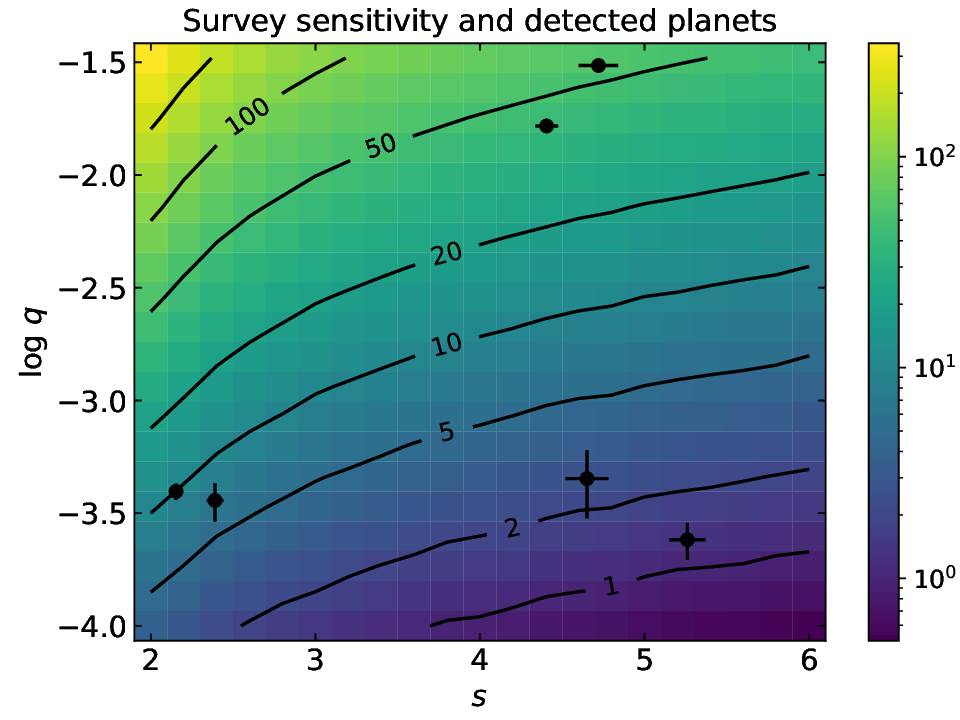} 
}
\FigCap{Survey detection efficiency as a function of separation and mass-ratio calculated using default detection criteria. Black symbols mark detected planets. Note that $(s, \log q) = (4.7, -1.5)$ planet OGLE-2016-BLG-0263Lb is detected only with extended selection criteria.
}
\end{figure}

We can combine the measured detection efficiency with the planet occurrence rate from Suzuki \etal (2016), denoted $f_\mathrm{S16}(s, q)$, to calculate the expected number of detections:
$$
N_\mathrm{exp,S16} = \int f_\mathrm{S16}(s, q)\, S(s, q)\, d\log s\,\, d\log q.
$$
For default (extended) selection criteria this integral results in 2.2 (2.4) planets and the probability of detecting as many planets as we detected or more, \ie $\geq 5$ ($\geq 6$), is $7.7\%$ ($3.6\%$) according to the Poisson distribution. These numbers indicate that the wide-orbit planet occurrence rate is higher than predicted by the extrapolation of the Suzuki \etal (2016) results.

\section{Wide-orbit Planet Occurrence Rate} 

We parameterize the planet occurrence rate as a function of $s$ and $q$:
$$
f(s, q; A, n, m) = \frac{d^2N_{pl}}{d\log q\,\,\,d\log s} = As^m\left(\frac{q}{q_\mathrm{br}}\right)^n. \eqno(6)
$$
We follow Suzuki \etal (2016) by fixing $q_\mathrm{br} \equiv 1.7\times10^{-4}$ so that our estimate of $A$ can be directly compared to their results. At the same time, we ignore a break in a slope of the mass-ratio function that Suzuki \etal (2016) found at $q_\mathrm{br}$. By ignoring this break we do not affect our results because we have very little sensitivity for $q<q_\mathrm{br}$ and we have not found any planet with such small $q$ value. Additionally, the value of $q$ at which the mass-ratio function breaks was re-investigated by Jung \etal (2019) who analyzed 15 microlensing planets with $q<3\times10^{-4}$ and estimated $q_\mathrm{br}$ of $0.55\times10^{-4}$, which is smaller than the lower limit of $q$ considered here.

To estimate the wide-orbit planet occurrence rate we use the Bayesian hierarchical inference, \ie we take into account the uncertainties of measured planet parameters (which are significant for some of the planets). In contrast, the previously used approach (used by, \eg Gould \etal 2010 and Suzuki \etal 2016) ignores the uncertainties of planet parameters. 

The number of planets detected is defined by the Poisson distribution, hence, under the assumption of negligible uncertainties of $s$ and $q$, the likelihood $\mathcal{L}_\mathrm{neg}$ is defined by (\eg Gould \etal 2010, Foreman-Mackey \etal 2014):
$$
\mathcal{L}_\mathrm{neg}(A, n, m) = e^{-N_\mathrm{exp}}\prod_{i=1}^{N_\mathrm{obs}}f(s_i, q_i;\,\, A, n, m)S(s_i, q_i) \eqno(7)
$$
$$
N_\mathrm{exp} = \int f(s, q;\,\, A, n, m)\, S(s, q)\, d\log s\,\, d\log q.
$$
We modify this formula following Foreman-Mackey \etal (2014). The last term in Equation~(7) uses point estimates of planet parameters and in the hierarchical approach we marginalize this term over posterior estimates for each planet. For this purpose, we draw $K$ samples $(s_{i,k}, q_{i,k})$ for each planet $i$. In our case, the posteriors of planet parameters were derived assuming flat priors on $q$ and $s$, hence, we can define hierarchical likelihood $\mathcal{L}$ as (Equation~(11) of Foreman-Mackey \etal 2014):
$$
\mathcal{L}(A, n, m) = e^{-N_\mathrm{exp}}\prod_{i=1}^{N_\mathrm{obs}} \left(\frac{1}{K}\sum_{k=1}^{K}f(s_{i,k}, q_{i,k};\, A, n, m)S(s_{i,k}, q_{i,k})\right).
$$

We obtain posterior distributions of $(A, n, m)$ by using the \textsc{EMCEE} package. As a prior, we use normal distributions with mean values taken from Suzuki \etal (2016) and standard deviations multiplied by a factor of 5.0. The resulting posterior distribution is presented in Table~2 (columns 2 and 3 for default and extended selection, respectively). In Table~2 we also present parameters derived assuming wide interpretation for OGLE-2017-BLG-0114 (which we consider unlikely; columns 4 and 5). The parameters derived by Suzuki \etal (2016) are given in the last column of Table~2 for comparison. We see that for the extended selection criteria $A$ and $n$ are very close to the Suzuki \etal (2016) values, contrary to the default selection criteria results. Parameter $m$ is larger than derived by Suzuki \etal (2016) in all cases, though its uncertainty is large. The last row of Table~2 provides the total number of planets per star in studied the range:
$$
N_\mathrm{tot} = \int^{\log 0.033}_{-4}\int^{\log 6}_{\log 2} f(s, q;\,\, A, n, m)\,\, d\log s\,\, d\log q. \eqno(8)
$$
All four our fits lead to $N_\mathrm{tot}$ significantly larger than calculated based on Suzuki \etal (2016). Our results indicate that the number of ice giant exoplanets is most probably larger than one per star. We present a corner plot for default selection in Figure~7. It reveals a significant correlation between $A$ and $m$. This correlation is seen also for extended selection.

Our anomaly detection method and planet parameter estimation are designed to detect a lens system composed of one star and one planet. We found high occurrence rate of wide-orbit planets which results in many stars having more than one wide-orbit planet, which may seem contradicting our assumptions. However, multiple-lens systems can be well approximated as a superposition of binary lenses (Han 2005) and we are analysing planetary caustic anomalies for which binary-lens approximation works in virtually all cases.

\MakeTable{l|r r|r r|r}{12.5cm}{
Planet occurrence rate parameters
}
{\hline
 &  default selection & extended selection & default selection & extended selection & Suzuki \etal  \\
 & 5 planets & 6 planets & 6 planets & 7 planets &  (2016) \\
\hline
$A$ & $ 1.04^{+0.78}_{-0.57} $ & $0.64^{+0.70}_{-0.41}$ & $ 0.87^{+0.77}_{-0.51} $ & $0.66^{+0.72}_{-0.42}$  & $0.61^{+0.21}_{-0.16}$ \\
$n$ & $ -1.15\pm0.25 $ & $-0.92\pm0.22$ & $ -1.04\pm0.24 $ & $-0.88\pm0.21$ & $-0.93\pm0.13$ \\
$m$ & $ 1.09\pm0.64 $ & $1.22\pm0.77$ & $ 1.16\pm0.70 $ & $1.25\pm0.73$ & $0.49^{+0.47}_{-0.49}$ \\
\hline
$N_\mathrm{tot}$ & $1.39^{+0.92}_{-0.59}$ & $1.12^{+0.75}_{-0.49}$ & $1.34^{+0.89}_{-0.58}$ & $1.24^{+0.78}_{-0.52}$ & $0.41^{+0.41}_{-0.21}$ $^a$ \\
\hline
\multicolumn{6}{p{12cm}}{For parameter definitions see Equations~(6) and (8). Columns 2 and 3 use planets from Table~1. Columns 4 and 5 additionally include the wide-orbit interpretation of OGLE-2017-BLG-0114 (see Section~4.4).} \\
\multicolumn{6}{p{12cm}}{$^a$ -- for consistency, this value neglects the change of $n$ for $q < q_\mathrm{br}$.}
}

\begin{figure}[htb]
{\centering
\includegraphics[width=\textwidth]{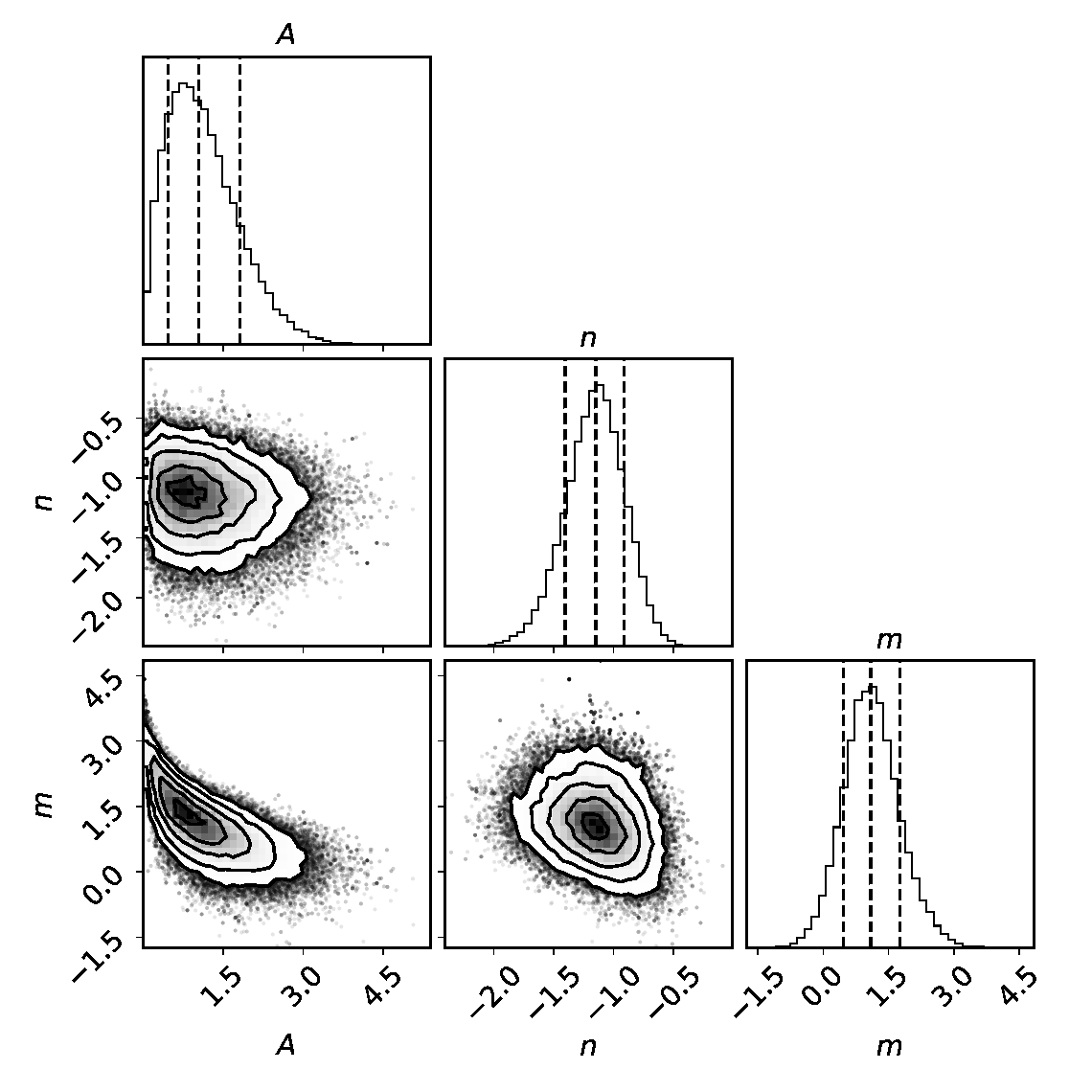}
}
\FigCap{Posterior distribution of planet occurrence rate parameters for three parameter fit with default selection criteria. Figure prepared using \textsc{Corner} code by Foreman-Mackey (2016).
}
\end{figure}

\subsection{Reliability of Detected Planets} 

Above, we have assumed that all events presented in Table~1 have in fact wide-orbit planet lenses. The reliability of the planetary interpretation for the events in our sample is unquestionable, but for OGLE-2011-BLG-0173 there was some ambiguity between $s>1$ and $s<1$ solutions. In Poleski \etal (2018) we used the prior planet rate (based on Suzuki \etal 2016) to compare posterior probabilities of $s=4.65$ and $s=0.22$ solutions and concluded that the wide solution is more likely. We can re-check this conclusion using planet rate derived here. For this purpose, we repeated the three parameter fit with default selection and with OGLE-2011-BLG-0173Lb removed from the sample of detected planets. The derived planet rate evaluated at $s=4.65$ and $q=4.5\times10^{-4}$ is higher by a factor of 1.9 than the Suzuki \etal (2016) results (and a factor of 3.5 if OGLE-2011-BLG-0173Lb is included). This makes even stronger argument for a wide-orbit planet interpretation than was presented in Poleski \etal (2018).

\section{Summary and Discussion} 

We performed an extensive search for wide-orbit planets in the OGLE-III and OGLE-IV data. This search revealed six known wide-orbit planets, three events which are definitely not wide-orbit planets, and a new anomalous event (OGLE-2017-BLG-0114) for which wide-orbit planet interpretation is not excluded but a different model is preferred. We also run detailed detection efficiency calculations, which included a thorough estimation of $\rho$. We presented conceptually simple and efficient method to measure the wide-orbit planet detection efficiency for parallax events. The resulting survey sensitivity is higher for large separations than in the previous microlensing studies. We combined the detected planets and detection efficiency to derive the wide-orbit planet occurrence rate. The rate is higher than derived previously by Suzuki \etal (2016) and we compare results in Table~2. We calculated the total number of planets per star in the studied range of $s$ and $q$ and for the default selection criteria it resulted in $1.4^{+0.9}_{-0.6}$. This value is higher by $2.4\sigma$ than the value calculated based on Suzuki \etal (2016) and shows that the wide-orbit planets are very common. In a separate paper, we will evaluate how this high occurrence rate of wide-orbit planets affects the interpretation of free-floating planet candidate events.

We also verified the wide-orbit planet interpretation for OGLE-2011-BLG-0173Lb. We found even stronger evidence than presented in a discovery paper.

There are wide-orbit microlensing planets that are not considered here because they either did not show the planetary caustic signal, were not detected in the OGLE data, or did not show an event caused by the host. We plot these planets (gray points) in Figure~8 together with planets found in this study (black points). The objects at the top of this diagram may be brown dwarfs. We see that the number of detected objects drops significantly at $s=3.5$: there are ten objects with $2<s<3.5$ and six with $3.5<s<7$. We also see that all except three objects are located on or above a diagonal line from $(s,\log q) = (2, -3.8)$ to $(5, -1.5)$. The other three objects are OGLE-2011-BLG-0173Lb, OGLE-2008-BLG-092LAb, and OGLE-2013-BLG-0911Lb. We see that when it comes to low mass-ratios, OGLE-2011-BLG-0173Lb and OGLE-2008-BLG-092LAb are the most extreme planets currently known.

\begin{figure}[htb]
{\centering
\includegraphics[width=\textwidth]{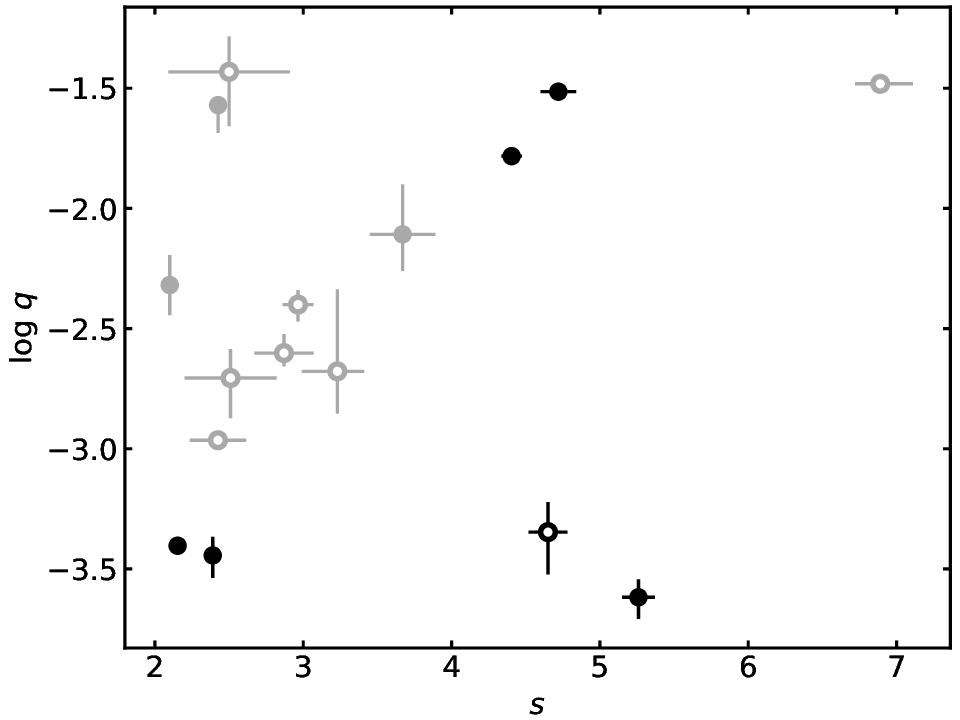}
}
\FigCap{Mass-ratio \vs separation diagram for all events with $s>2$ and $q<0.033$. Filled and open circles mark planets not affected and affected by the close-wide degeneracy, respectively. Black points indicate planets from Table~1, \ie are the same as points in Figure~6. Gray points mark planets not detected in this study or with close solution being more likely in order of increasing $s$: MOA-bin-1Lb (Bennett \etal 2012), OGLE-2012-BLG-0563Lb (Fukui \etal 2015), OGLE-2014-BLG-1112Lb (Han \etal 2017$b$), KMT-2019-BLG-1339Lb (Han \etal 2020$c$), KMT-2019-BLG-1953Lb (Han \etal 2020$b$), MOA-2007-BLG-400Lb (Dong \etal 2009), KMT-2016-BLG-1107Lb (Hwang \etal 2019), OGLE-2017-BLG-0114Lb (this paper), KMT-2016-BLG-1107Lb (Hwang \etal 2019), OGLE-2016-BLG-1227Lb (Han \etal 2020$a$), and OGLE-2013-BLG-0911Lb (Miyazaki \etal 2020).
}
\end{figure}

We give details of the observed fields, events detected, survey detection efficiency, and posterior sampling at the OGLE Internet archive:
\begin{center}
\textit{http://www.astrouw.edu.pl/ogle/ogle4/planets/wide\_orbit\_rate/}
\textit{ftp://ftp.astrouw.edu.pl/ogle/ogle4/planets/wide\_orbit\_rate/}
\end{center}

\Acknow{
We would like to thank I.~Soszy\'nski, M.~Kubiak, G.~Pie\-trzy\-\'nski, {\L}.~Wyrzykowski, M.~Pawlak, and S.~Koz{\l}owski for their contribution to the collection of the OGLE photometric data. 
Work by RP was supported by Polish National Agency for Academic Exchange grant ``Polish Returns 2019.'' 
The OGLE project has received funding from the National Science Centre, Poland, grant MAESTRO 2014/14/A/ST9/00121 to AU. 
}

\end{document}